\documentclass[english,reprint,amsmath,amssymb,aps,superscriptaddress,floatfix]{revtex4-1}
\usepackage[T1]{fontenc}
\usepackage[latin9]{inputenc}
\usepackage{color}
\usepackage{babel}
\usepackage{amsmath}
\usepackage{graphicx}
\usepackage{esint}
\usepackage{multirow}
\usepackage{bm}
\usepackage[caption=false]{subfig}
\usepackage[unicode=true,pdfusetitle,
 bookmarks=true,bookmarksnumbered=false,bookmarksopen=false,
 breaklinks=false,pdfborder={0 0 1},backref=false,colorlinks=true]
 {hyperref}
\hypersetup{
 linkcolor={blue},citecolor={blue},urlcolor={blue}}

\makeatletter

\@ifundefined{textcolor}{}
{%
 \definecolor{BLACK}{gray}{0}
 \definecolor{WHITE}{gray}{1}
 \definecolor{RED}{rgb}{1,0,0}
 \definecolor{GREEN}{rgb}{0,1,0}
 \definecolor{BLUE}{rgb}{0,0,1}
 \definecolor{CYAN}{cmyk}{1,0,0,0}
 \definecolor{MAGENTA}{cmyk}{0,1,0,0}
 \definecolor{YELLOW}{cmyk}{0,0,1,0}
}

\@ifundefined{showcaptionsetup}{}{%
 \PassOptionsToPackage{caption=false}{subfig}}
\makeatother

\begin{document}

\title{Hierarchy of Information Scrambling, \\ Thermalization, and
Hydrodynamic Flow in Graphene}

\author{Markus J. Klug }

\affiliation{Institut f\"ur Theorie der Kondensierten Materie, Karlsruher Institut
f\"ur Technologie, 76131 Karlsruhe, Germany}

\author{Mathias S. Scheurer }

\affiliation{Institut f\"ur Theorie der Kondensierten Materie, Karlsruher Institut
f\"ur Technologie, 76131 Karlsruhe, Germany}

\affiliation{Department of Physics, Harvard University, Cambridge MA 02138, USA}

\author{J\"{o}rg Schmalian}

\affiliation{Institut f\"ur Theorie der Kondensierten Materie, Karlsruher Institut
f\"ur Technologie, 76131 Karlsruhe, Germany}

\affiliation{Institut f\"ur Festk\"orperphysik, Karlsruher Institut f\"ur Technologie,
76344 Karlsruhe, Germany}

\date{\today }
\begin{abstract}
We determine the information scrambling rate $\lambda_{L}$ due to
electron-electron Coulomb interaction in graphene. $\lambda_{L}$
characterizes the growth of chaos and has been argued to give information
about the thermalization and hydrodynamic transport coefficients of
a many-body system. We demonstrate that $\lambda_{L}$ behaves for
strong coupling similar to transport and energy relaxation rates.
A weak coupling analysis, however, reveals that scrambling is related
to dephasing or single particle relaxation. Furthermore, $\lambda_{L}$ is found to be parametrically larger than the collision rate relevant for hydrodynamic
processes, such as electrical conduction or viscous flow, and the rate of energy relaxation, relevant for thermalization. 
Thus, while scrambling is obviously necessary for thermalization and quantum transport, it does generically not set the time scale for these processes.
In addition we
derive a quantum kinetic theory for information scrambling that resembles
the celebrated Boltzmann equation and offers a physically transparent
insight into quantum chaos in many-body systems. 
\end{abstract}

\maketitle
The emergence of chaos is the most plausible explanation for the thermalization
of closed quantum many-body systems. An efficient concept to quantify
quantum chaos is through the scrambling rate $\lambda_{L}$ \cite{Larkin1969,Sekino2008}.
After the time $\lambda_{L}^{-1}$ of the evolution of an out-of-equilibrium
initial state, quantum entanglement has spread across the system.
Then, the initial state cannot be recovered, i.e. unscrambled, via
local measurements. The system has lost its memory, a key prerequisite
for thermalization. Still, it is unclear whether the scrambling rate
is indeed the characteristic scale that determines the return to thermal
equilibrium. Similar to the scrambling rate with respect to temporal
evolution, the butterfly velocity $v_{B}$ characterizes the corresponding
spread of entanglement in space after a local perturbation. Formally,
these quantities can be determined from the growth in time or space
of commutators or anticommutators of local operators $A$ and $B$,  
\begin{equation}
\big\langle\left|\left[A\left(\mathbf{x},t\right),B\left(\mathbf{0},0\right)\right]\right|^{2}\big\rangle\sim e^{2\lambda_{L}\left(t-\frac{|\mathbf{x}|}{v_{B}}\right)}.
\end{equation}

\begin{figure}[b]
\centering{}\includegraphics[width=.96\columnwidth]{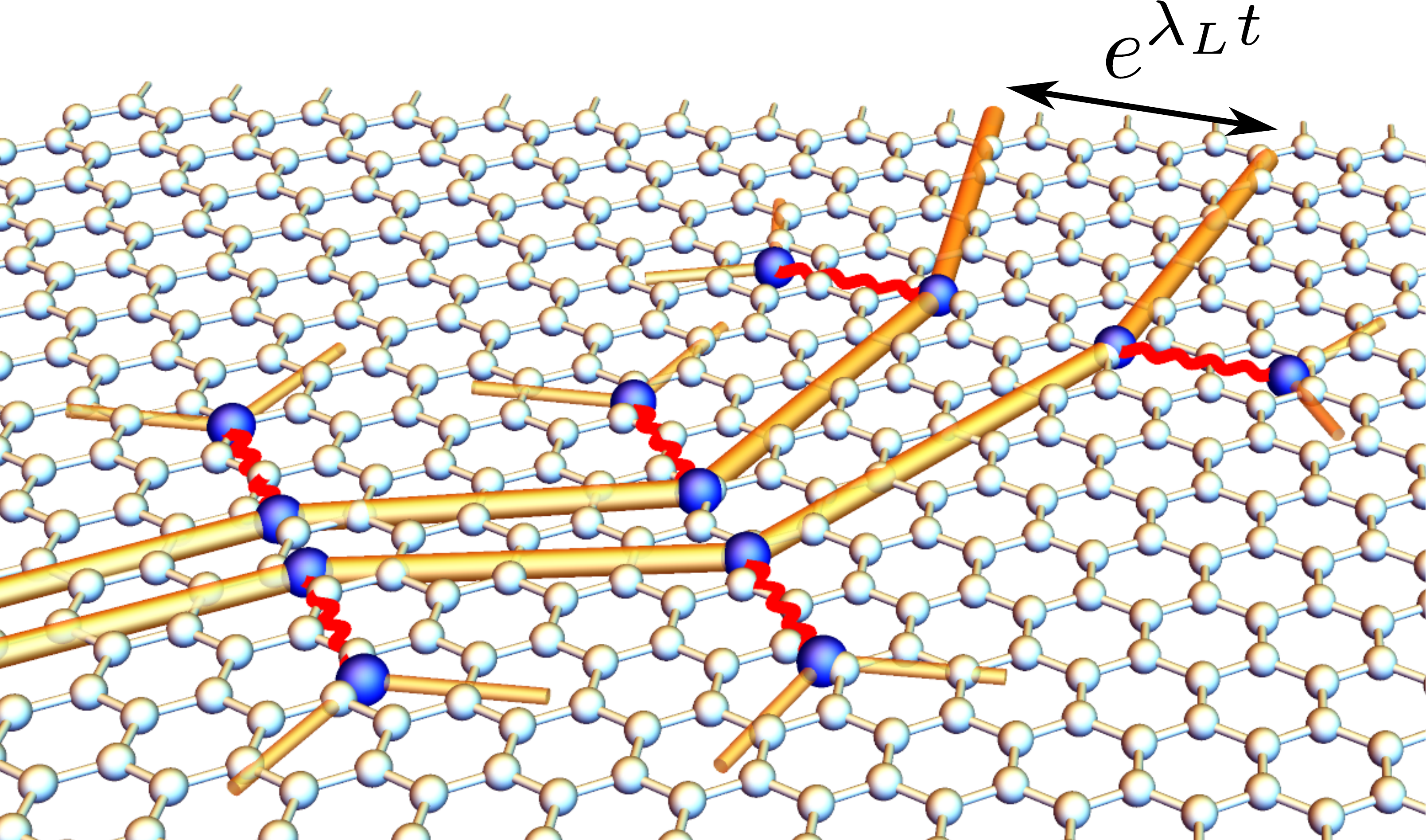}\protect\caption{\label{fig:Quantum-chaos-in}Exponentially diverging electron trajectories
(orange) in graphene resulting from Coulomb interaction (red).}
\end{figure}

A transparent interpretation of this squared commutator exists in
the quasi-classical limit for the motion of a particle with initial
coordinate $q\left(0\right)$ and conjugate momentum $p\left(t\right)$
for which $\langle[p\left(t\right),p\left(0\right)]^{2}\rangle=\langle\big(\frac{\partial p\left(t\right)}{\partial q\left(0\right)}\big)^{2}\rangle$\cite{Larkin1969}.
Exponential growth behavior of this correlator is e.g. realized by
electrons in a weakly disordered metal \cite{Syzranov17}. Thus, the
expectation value measures the dependence of the momentum at time
$t$ as one changes the initial coordinates, a key measure for chaotic
motion. The square in the commutator ensures that positive and negative
momenta at time $t$ do not average to zero. The corresponding spread
of an initial state of two nearby electrons in graphene is sketched
in Fig. \ref{fig:Quantum-chaos-in}.

While the formal interpretation of scrambling is established in the
information theoretic sense, with key applications in quantum circuits
\cite{Hayden2007}, it is not obvious how to measure $\lambda_{L}$
or $v_{B}$ for a generic solid-state system. In fact it is unclear
which specific physical observable might essentially probe these quantities
and how this relates to thermalization. For example a close link to
transport quantities was proposed and $v_{B}^{2}/\lambda_{L}$ related
to the charge \cite{Blake2016,Blake2016_2} or heat \cite{Patel2017}
diffusivities. The relation between scrambling and transport seems
consistent with the bound on chaos $\lambda_{L}\leq2\pi k_{B}T/\hbar$,
derived under rather general conditions \cite{Maldacena2016}, a bound
that is saturated by the Sachdev-Ye-Kitaev model \cite{Sachdev1993,Kitaev2015,Maldacena16}
and more generally in models with holographic duals. Together with
the Planckian transport rate $\tau_{tr}^{-1}\approx k_{{\rm B}}T/\hbar$
that emerges in some strongly correlated systems \cite{Zaanen2004,Sachdev2006},
this suggests a connection between scrambling and transport processes.
On the other hand, the analysis of weakly interacting diffusive electrons
revealed that $\lambda_{L}$ is rather determined by the single particle
scattering rate \cite{Patel2017_2}. In addition, the scrambling rate
of a bad metal, coupled to long-lived phonons was shown to be determined
by phonons and not by the short transport time \cite{Werman2017}.

\begin{figure}[t]
\centering{}\includegraphics[width=1\columnwidth]{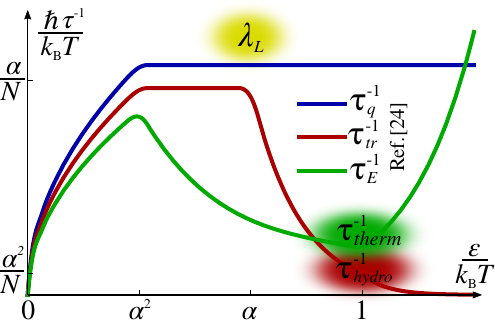}\protect\caption{\label{fig:Qualitative-representation-the}Qualitative representation of the energy dependence
of the scattering rates, the dephasing rate $\tau_{q}^{-1}$, the rate
relevant for transport processes $\tau_{tr}^{-1}$ and the energy
relaxation rate $\tau_{E}^{-1}$, for weak coupling, $\alpha \ll 1$, obtained in
Ref. \cite{Schuett2011}. Rates, which are relevant to hydrodynamic
transport and energy relaxation, are determined by excitation energies $\epsilon\sim k_{B}T$
and are represented by a red or green spot. The determined scrambling rate
$\lambda_{L}$ for weak coupling probing excitations predominantly
at energies $\epsilon\sim\alpha k_{B}T$, is represented by a yellow
spot. Scrambling processes at weak coupling are therefore a lot faster
than the collisions that give rise to the hydrodynamic behavior of
graphene, and also faster than the energy relaxation rate that one
would expect to govern thermalization. Instead, the scrambling time
in graphene is closely related to the dephasing scattering rate. }
\end{figure}

Graphene is a unique condensed matter system owed to its Dirac spectrum.
On the one hand, recent experiments demonstrated hydrodynamic flow
of the electron fluid with giant magneto-drag \cite{Titov2013}, a
breakdown of the Wiedemann-Franz law \cite{Crossno}, super-ballistic
transport \cite{Guo,KrishnaKumar}, and negative local resistance
\cite{Bandurin2016,Levitov2016}. Even nonlinear phenomena, like hydrodynamic
hot spot relaxation from out-of-equilibrium configurations have been
proposed \cite{Briscot,Naroshny2017}. Rapid local thermalization
is crucial for the applicability of hydrodynamic descriptions. On
the other hand, graphene, like weakly interacting diffusive electrons
discussed in \cite{Patel2017_2}, displays a number of distinct scattering
rates -- for single particle excitations $\tau_{q}^{-1}\left(\epsilon\right)$,
for energy relaxations $\tau_{E}^{-1}\left(\epsilon\right)$, or for
transport processes $\tau_{{\rm tr}}^{-1}\left(\epsilon\right)$;
see Ref. \cite{Schuett2011}. In Fig. \ref{fig:Qualitative-representation-the}
we plot these characteristic time scales as function of the particle energy $\epsilon$. While
the single-particle rate is always the largest, it depends on the
characteristic energy whether the rate of energy relaxation or the
transport rate is larger. For a summary of these time scales see
also Appendix \ref{sec:Characteristic-Rates}. The origin for these
distinct scales is the infrared-singular collision kernel. It allows
to identify to what extend scrambling and hydrodynamic collisions
are related. In addition, it allows to distinguish thermalization,
that should be governed by the energy relaxation rate at $\epsilon\sim k_{B}T$,
from scrambling. 

In the first part of this work, we determine the scrambling rate $\lambda_{L}$
for electrons in graphene with electron-electron Coulomb interaction
using a diagrammatic approach, presented e.g.~in Ref.~\cite{Patel2017,Patel2017_2,Werman2017}.
Details of the considered microscopic model are presented in Sec.
\ref{sec:The-Model}, whereas the computation of $\lambda_{L}$ is
found in Sec. \ref{sec:The-out-of-time-order-correlator}. The analysis
is done for large $N$, where $N$ is the number of fermion flavors
($N=4$ in graphene). This allows the determination of $\lambda_{L}$
for arbitrary values of the effective fine structure constant $\alpha=Ne^{2}/\hbar v$
of graphene, which is formally considered to be $N$-independent \cite{Son2007}.
There $v$ and $e$ denote the bare Fermi velocity and the electron
charge, respectively. At strong coupling we find 
\begin{equation}
\lambda_{L}\left(\alpha\gg1\right)\cong0.80\,\frac{4}{N}\frac{2\pi k_{{\rm B}}T}{\hbar},\label{eq:lambda-strong}
\end{equation}
a behavior that is consistent with other large-$N$ calculations \cite{Chowdhury2017}.
Due to the large-$N$ expansion this result is parametrically far
from the above bound. However, it does behave like the transport relaxation
rate $\tau_{{\rm tr}}^{-1}(k_B T)\sim k_{B}T/\hbar N$ that occurs in the
same limit \cite{Link2017}. For large $\alpha$ the only characteristic
scale is $k_{B}T$ making a clear association of $\lambda_{L}$with
a specific time scale difficult. Therefore, our analysis is more revealing
in the weak coupling limit, where we find 
\begin{equation}
\lambda_{L}\left(\alpha\ll1\right)\gtrsim0.37\,\alpha\frac{4}{N}\frac{2\pi k_{B}T}{\hbar}.\label{eq:lambda-weak}
\end{equation}
This result is parametrically larger than the transport rate $\tau_{{\rm tr}}^{-1}(k_B T)\sim\alpha^{2}k_{B}T/\hbar N$
that occurs in the d.c.~conductivity \cite{Fritz2008,Kashuba2008}
or the electron viscosity \cite{Mueller2009}. Scrambling processes
at weak coupling are therefore a lot faster than the collisions that
give rise to the hydrodynamic behavior of graphene. $\lambda_{L}$
is also faster than the energy relaxation rate $\tau_{E}^{-1}(k_B T)\sim\alpha^{2}\log(\alpha^{-1})k_{B}T/\hbar N$
that one would expect to govern thermalization. Thus, scrambling and
thermalization are two clearly distinct phenomena. Instead, the scrambling
time in graphene is closely related to the quantum or dephasing scattering
rate \cite{Schuett2011} $\tau_{q}^{-1}\left(\epsilon\right)$ for
characteristic energies $\epsilon\sim\alpha k_{B}T$ that are determined
by the screening length $l_{s}^{-1}\approx\ln2\alpha k_{B}T/v\hbar$.
This behavior is illustrated in Fig. \ref{fig:Qualitative-representation-the}. 

In order to obtain a clear physical understanding of information scrambling
in many-body systems, we also present an alternative approach to quantum
chaos using non-equilibrium techniques in Sec. \ref{sec:Kinetic-equation}, similar in spirit to the methods presented in Ref. \cite{Aleiner2016,Grozdanov2018}. We derive a kinetic equation similar to the Boltzmann equation in
form of an integro-differential equation describing the growth and
spread of a small, localized perturbation. It is shown explicitly that this approach
reproduces the results obtained within the diagrammatic formalism.

\section{The model\label{sec:The-Model}}

We consider the following Hamiltonian for electrons in graphene near
the Dirac point with electron-electron Coulomb interaction (setting
$\hbar=1$) 
\begin{align}
H & =v\sum_{i=1}^{N}\int d^{2}\mathbf{x}\psi_{i}^{\dagger}\left(\mathbf{x}\right)\frac{1}{i}\mathbf{\nabla}\cdot\mathbf{\boldsymbol{\sigma}}\psi_{i}\left(\mathbf{x}\right)\label{eq:H0}\\
+ & \frac{e^{2}}{2}\sum_{\substack{ij=1\\
\alpha\beta
}
}^{N}\int d^{2}\mathbf{x}d^{2}\mathbf{x}'\frac{\psi_{i\alpha}^{\dagger}\left(\mathbf{x}\right)\psi_{j\beta}^{\dagger}\left(\mathbf{x}'\right)\psi_{j\beta}\left(\mathbf{x}'\right)\psi_{i\alpha}\left(\mathbf{x}\right)}{\left|\mathbf{x}-\mathbf{x}'\right|}.\nonumber 
\end{align}
Here, $\psi_{i}=(\psi_{i1},\psi_{i2})^{\text{T}}$ is a two component
spinor and $\mathbf{\boldsymbol{\sigma}}=(\sigma_{x},\sigma_{y})^{\text{T}}$
are Pauli matrices acting in pseudo-spin space. $i=1\dots N$ is an
additional flavor index that includes spin and valley degrees of freedom.
While $N=4$ for graphene, we keep $N$ arbitrary to be able to perform
a controlled expansion in $1/N$ \cite{Son2007,Foster2008}. A key
justification to use this approach for the description of graphene
comes from experiment. Several measurements clearly reveal interactions
effects \cite{Elias2011,Siegel2011,Yu2013}, but of a kind that is
fully consistent with renormalization group assisted perturbation
theory \cite{Sheehy2007}.

The retarded fermionic propagator, on the bare level given by 
\begin{equation}
(G^{R})^{-1}=\left(\omega+i0^{+}\right)\sigma^{0}-v\mathbf{k}\cdot\boldsymbol{\sigma},
\end{equation}
is dressed in leading order in $1/N$ by the usual rainbow diagram
for the retarded self energy\emph{ 
\begin{equation}
\Sigma_{\alpha\beta}^{R}\left(\mathbf{k},\omega\right)=\frac{i}{2N}(D^{R}\circ G_{\alpha\beta}^{K}+D^{K}\circ G_{\alpha\beta}^{K})_{\left(\mathbf{k},\omega\right)},
\end{equation}
}where the superscripts $R$, $K$, and later $A$ stand for retarded,
Keldysh, and advanced components of the Green's functions, and $\circ$
stands for a convolution with regards to frequencies and momenta.
$D^{-1}=D_{0}^{-1}+\Pi$ is the collective plasmon propagator with
bare Coulomb interaction $D_{0}^{R}=\frac{2\pi e^{2}}{\left|\mathbf{q}\right|}$
and the bosonic self energy 
\begin{equation}
\Pi^{R}\left(\mathbf{k},\omega\right)=\frac{i}{2}(G_{\alpha\beta}^{R}\circ G_{\beta\alpha}^{K}+G_{\alpha\beta}^{K}\circ G_{\beta\alpha}^{A})_{\left(\mathbf{k},\omega\right)}
\end{equation}
 which is of order $N^{0}$.

\section{Diagrammatic Approach\label{sec:The-out-of-time-order-correlator}}

We start from the 'regularized' version of the squared anticommutator
\begin{multline}
f_{\gamma\delta}^{\alpha\beta}\left(x_{1},x_{2},x_{3},x_{4}\right)=\frac{1}{N^{2}}\sum_{i,j=1}^{N}{\rm tr}\left(\left\{ \psi_{i\alpha}\left(x_{1}\right),\psi_{j\beta}^{\dagger}\left(x_{2}\right)\right\} \right.\\
\times\left.\sqrt{\rho}\left\{ \psi_{j\delta}\left(x_{4}\right),\psi_{i\gamma}^{\dagger}\left(x_{3}\right)\right\} \sqrt{\rho}\right)\label{eq:f(x)}
\end{multline}
where the regularization amounts to splitting the density matrix $\rho=Z^{-1}e^{-H/k_{B}T}$
between the two anticommutators \cite{Shenker2014,Maldacena2016}.
Otherwise the exponent would depend explicitly on the UV cut-off of
our effective field theory and would therefore be ill defined. $x=\left(\mathbf{x},t\right)$
stands for space and time coordinates. In order to determine the scrambling
time, we analyze 
\begin{equation}
C\left(t\right)=\theta\left(t\right)\int d^{2}\mathbf{x}\sum_{\alpha\beta}f_{\alpha\beta}^{\alpha\beta}\left(x,0,x,0\right),\label{eq:C(t)}
\end{equation}
which contains a correlator with non-trivial time order denoted out-of-time-order
correlator (OTOC), i.e.~sequences of operators which cannot be represented
on a conventional Keldysh contour evolving back and forth in time.

\subsection{Out-of-time-order formalism}

\begin{figure}[b]
\centering{} \includegraphics[width=1\columnwidth]{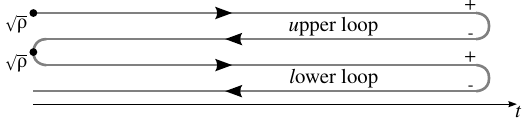}
\protect\caption{\label{fig:Keldysh-loop}Two-loop Keldysh contour. Index $u$ and
$l$ refer to the fields residing on the upper and lower Keldysh loop,
index $+$ and $-$ to fields residing on the forward and backward
propagating branch, respectively. The square roots of density matrices
are placed at the left turning points of the contour denoted by solid
black dots. }
\end{figure}

To determine the function $f$, we use the out-of-time-order formalism
and the corresponding four-branch (two-loop) Keldysh contour \cite{Aleiner2016,Stanford2016}. 
In this case,
the thermal expectation value is expressed as expectation value of
four-component Grassmann fields $\psi_{i\alpha}(t)=\left(\psi^{u+},\psi^{u-},\psi^{l+},\psi^{u-}\right)_{i\alpha,\left(t\right)}^{\text{T}}$,
and analogously $\bar{\psi}_{i\alpha}$, placed on the Keldysh contour
according to their relative causal position. The position of the fields
is specified, besides the time parameter $t$, by the loop index $\sigma=\left\{ u,l\right\} $
for the upper and lower loop, and the branch index $\left\{ +,-\right\} $
denoting the branches propagating forward and backward in time, respectively.
The contour including the position of the fractions of density matrices
placed between the anticommutators is depicted in Fig. \ref{fig:Keldysh-loop}.

We perform the standard Keldysh rotation \cite{Kamenev2011} for each
of the two loops $\sigma=\{u,l\}$ separately, 
\begin{align}
\psi^{\sigma\text{cl}} & =\frac{1}{\sqrt{2}}\left(\psi^{\sigma+}+\psi^{\sigma-}\right),\,\psi^{\sigma\text{q}}=\frac{1}{\sqrt{2}}\left(\psi^{\sigma+}-\psi^{\sigma-}\right),\notag\\
\bar{\psi}^{\sigma\text{cl}} & =\frac{1}{\sqrt{2}}\left(\bar{\psi}^{\sigma+}-\bar{\psi}^{\sigma-}\right),\,\bar{\psi}^{\sigma\text{q}}=\frac{1}{\sqrt{2}}\left(\bar{\psi}^{\sigma+}+\bar{\psi}^{\sigma-}\right),
\end{align}
where Keldysh indices $s=\left\{ \text{cl},\text{q}\right\} $ denote
'classical'- (cl) and 'quantum'- (q) field components, respectively. 

An effective Keldysh action is obtained by introducing real plasmon
fields $\phi$ which couple linearly to a pair of fermion fields.
For this, we carry out the standard Hubbard-Stratonovich transformation
to decouple the interaction term of the Hamilton operator in Eq.~\ref{eq:H0} in the charge
channel. Consequently, the quadratic part of the Keldysh action is
given by (pseudo-spin and flavor indices are dropped): 
\begin{multline}
S_{0}\left[\psi,\bar{\psi},\phi\right]=\sum_{\sigma^{(')}s^{(')}}\bigg(\int_{k}\bar{\psi}_{k}^{\sigma s}\, G_{\sigma\sigma',ss'}^{-1}(k)\,\psi_{k}^{\sigma's'}+\\
+\int_{q}\phi_{-q}^{\sigma s}\, D_{\sigma\sigma',ss'}^{-1}(q)\,\phi_{q}^{\sigma's'}\bigg),
\end{multline}
where $k=\left(\epsilon,\mathbf{k}\right)$ and $\int_{k}=\int\frac{d\epsilon}{2\pi}\int\frac{d^{2}\mathbf{k}}{\left(2\pi\right)^{2}}$.
The intra-loop components of the fermionic and bosonic plasmon propagators
($\sigma=\sigma'$) have the usual causal structure 
\begin{equation}
G_{\sigma\sigma,ss'}=\begin{pmatrix}G^{R} & G^{K}\\
0 & G^{A}
\end{pmatrix}_{ss'}\text{, }D_{\sigma\sigma,ss'}=\begin{pmatrix}D^{K} & D^{R}\\
D^{A} & 0
\end{pmatrix}_{ss'}
\end{equation}
where superscripts $R$/$A$ denote retarded and advanced components,
and the fermionic and bosonic Keldysh components are given by $G^{K}\left(k\right)=2i\tanh(\frac{\epsilon}{2k_{B}T})\text{Im}G^{R}\left(k\right)$
and $D^{K}\left(k\right)=2i\coth(\frac{\epsilon}{2k_{B}T})\text{Im}D^{R}\left(k\right)$.
In the case of inter-loop correlations ($\sigma'=\bar{\sigma}$ where
$\bar{u}=l$ and vice versa), 
\begin{equation}
G_{\sigma\bar{\sigma},ss'}=\begin{pmatrix}0 & G_{\sigma\bar{\sigma}}^{K}\\
0 & 0
\end{pmatrix}_{ss'}\text{, }D_{\sigma\bar{\sigma},ss'}=\begin{pmatrix}D_{\sigma\bar{\sigma}}^{K} & 0\\
0 & 0
\end{pmatrix}_{ss'},\label{eq:inter-loop_comp}
\end{equation}
there is only a Keldysh components which relates to the retarded components
as \begin{subequations} \label{eq:inter-loop-components} 
\begin{align}
D_{ul}^{K}(\omega,\mathbf{q}) & =\frac{2i\text{Im}D^{R}(\omega,\mathbf{q})}{\sinh\left(\frac{\omega}{2k_{B}T}\right)},\\
G_{lu,a}^{K}\left(\epsilon,\mathbf{k}\right) & =\frac{2i\text{Im}G_{a}^{R}(\epsilon,\mathbf{k})}{\cosh\left(\frac{\epsilon}{2k_{B}T}\right)}=-G_{ul,a}^{K}\left(\epsilon,\mathbf{k}\right).\label{eq:fermionic-inter-loop-components}
\end{align}
\end{subequations}Here, we use the band basis where the first term
in Eq. \ref{eq:H0} is diagonal and $G_{a}^{R}=\left(\omega+i0^{+}-av|\mathbf{k}|\right)^{-1}$
with $a=\pm1$.

The fermionic and bosonic fields couple via the term 
\begin{equation}
S_{\text{int}}\left[\psi,\bar{\psi},\phi\right]=\frac{1}{\sqrt{N}}\sum_{\sigma s^{(','')}}\int_{kq}\bar{\psi}_{k+q}^{\sigma s}\psi_{k}^{\sigma s'}\phi_{q}^{\sigma s''}\,\gamma_{ss'}^{s''}.
\end{equation}
which is weighted by a factor of $1/\sqrt{N}$. The coupling is diagonal
in Keldysh-loops as well as all dropped indices. The coupling vertices
acting in Keldysh component space are $\gamma_{ss'}^{\text{cl}}=\delta_{ss'}$
and $\gamma_{ss'}^{\text{q}}=\left(\sigma_{1}\right)_{ss'}$ with
$\sigma_{1}$ being the first Pauli matrix. 

Within this framework, the squared anticommutator Eq. \ref{eq:f(x)}
is recast as the expectation value of 'classical'- and 'quantum'-Keldysh
field components 
\begin{multline}
f_{\gamma\delta}^{\alpha\beta}\left(x_{1},x_{2},x_{3},x_{4}\right)=\\
-\frac{1}{N^{2}} \sum_{i,j=1}^{N}\langle\psi_{i\alpha}^{l\text{cl}}\left(x_{1}\right)\bar{\psi}_{j\beta}^{l\text{cl}}\left(x_{2}\right)\bar{\psi}_{i\gamma}^{u\text{q}}\left(x_{3}\right)\psi_{j\delta}^{u\text{q}}\left(x_{4}\right)\rangle_{\mathcal{K}}
\end{multline}
which are evaluated with respect to the two-loop Keldysh action $\langle\dots\rangle_{\mathcal{K}}=\int\mathcal{D}\left(\bar{\psi},\psi,\phi\right)\dots e^{i\mathcal{S}_{\mathcal{K}}\left[\psi,\phi\right]}$
with $S_{\mathcal{K}}=S_{0}+S_{\text{int}}$. Eventually, contributions
of $S_{\text{int }}$ to $f$ are incorporated perturbatively in orders
of $1/N$.

\subsection{Bethe-Salpeter equation\label{sub:Bethe-Salpeter-equation}}

\begin{figure}[t]
\subfloat[\label{fig:diagrams-scrambling-a}]{\centering{}\includegraphics[height=0.2\columnwidth]{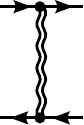}}\qquad{}\subfloat[\label{fig:diagrams-scrambling-b}]{\centering{}\includegraphics[height=0.2\columnwidth]{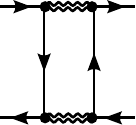}

}\qquad{}\subfloat[]{\centering{}\includegraphics[height=0.2\columnwidth]{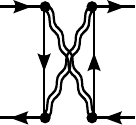}

}

\subfloat[\label{fig:Diagrammatic-representation-of-f}]{\centering{}\includegraphics[height=0.2\columnwidth]{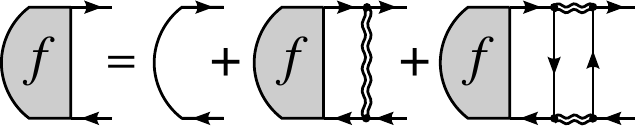}

}

\protect\caption{\label{fig:diagrams-scrambling}Diagrammatic representation of processes
contributing to scrambling in leading order in $1/N$. Straight lines
correspond to bare fermion propagators, wavy lines to plasmon propagators
dressed by particle-hole fluctuations. Both types of propagators are
of order $N^{0}$. Each interaction vertex contributes a factor of
$1/\sqrt{N}$, whereas each fermion loop a factor of $N$. Propagators
residing on the upper or lower horizontal line are located on one
of the Keldysh loops, vertical propagators are loop-connecting propagators
causing scrambling. (a) One-rung diagram containing one inter-loop
bosonic propagator. (b) Two-rung diagram containing two inter-loop
fermionic propagators. An extra factor of $N$ is obtained by an additional
closed fermion loop and is therefore of order $\mathcal{O}(1/N)$.
The crossed-rung diagram in (c) vanishes exactly as the retarded and
advanced inter-loop components (see Eq.~\ref{eq:inter-loop_comp})
are zero. (d) Diagrammatic representation of self-consistent Bethe-Salpeter 
equation for $f(\omega,k)$ in leading order in $1/N$. }
\end{figure}

Interaction processes contributing to scrambling involve inter-loop
correlators. The leading order processes in $1/N$ are depicted in
Fig. \ref{fig:diagrams-scrambling-a} and \ref{fig:diagrams-scrambling-b}
as diagrammatic representation. To capture the exponential growth
behavior of Eq.~\ref{eq:C(t)}, these diagrams are summed in an infinite
ladder series. Note that irreducible contributions which are of higher
order in $1/N$ yield corrections to the growth rate $\lambda_{L}$
of higher order and are therefore neglected. 
The diagrammatic series is derived by considering the Laplace transform
of Eq.~\ref{eq:f(x)} where half of the internal degrees of freedom
are already traced out: 
\begin{align}\begin{split}
f_{\beta}^{\alpha}(\omega,k)&\equiv\int_{\epsilon'\mathbf{k'}}\int_{x_{1}x_{2}x_{3}}\sum_{\gamma}f_{\beta\gamma}^{\alpha\gamma}(x_{1},x_{2},x_{3},0)\\
\times &e^{i\left(\omega+\epsilon'\right)t_{1}-i\left(\omega+\epsilon\right)t_{2}-i\epsilon t_{3}-i\mathbf{k}\cdot\left(\mathbf{x}_{1}-\mathbf{x}_{3}\right)+i\mathbf{k}'\cdot\mathbf{x}_{2}}\label{eq:f-w-k} \\
&\equiv  \,\, \vcenter{\hbox{\includegraphics[height=4.2em]{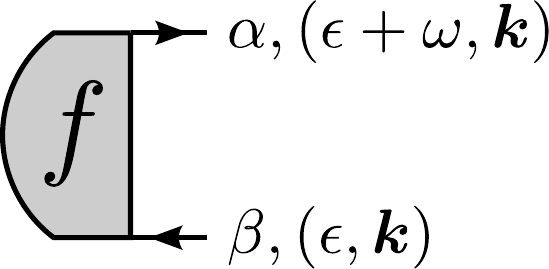}}}\end{split}
\end{align}
with $k=\left(\epsilon,\mathbf{k}\right)$, $\int_{k}=\int\frac{d\epsilon}{2\pi}\int\frac{d^{2}\mathbf{k}}{\left(2\pi\right)^{2}}$
and $\int_{x}=\int dt\int d^{2}\mathbf{x}$. The resulting Bethe-Salpeter
equation, which determines $f$ recursively, is given by 
\begin{gather}
f_{\beta}^{\alpha}\left(\omega,k\right)=\frac{1}{N}\sum_\gamma G_{\alpha\gamma}^{R}(\omega+\epsilon,\mathbf{k})G_{\gamma\beta}^{A}(k)+\qquad\qquad\quad\label{eq:bse}\\
\frac{1}{N}\int_{k'}\sum_{\gamma^{(')}\delta^{(')}}f_{\delta'}^{\gamma'}(\omega,k')\Gamma_{\delta\delta'}^{\gamma\gamma'}\left(\omega,k,k'\right)G_{\alpha\gamma}^{R}(\omega+\epsilon,\mathbf{k})G_{\delta\beta}^{A}(k)\nonumber 
\end{gather}
and depicted in Fig.~\ref{fig:Diagrammatic-representation-of-f} as
diagrammatic representation. The inter-loop scattering vertex $\Gamma$
contains one-rung (first term) and two-rung (second term) contributions,
\begin{multline}
\Gamma_{\gamma\delta}^{\alpha\beta}\left(\omega,k,k'\right)=i\delta_{\alpha\beta}\delta_{\gamma\delta}D_{ul}^{K}(k-k')\\
+\int_{\tilde{k}}G_{\alpha\gamma}^{lu}(k-\tilde{k})G_{\delta\beta}^{ul}(k'-\tilde{k})D^{R}(\omega+\tilde{\epsilon},\tilde{\mathbf{k}})D^{A}(\tilde{k}),\label{eq:scattering-vertex}
\end{multline}
shown in Fig.~\ref{fig:diagrams-scrambling}(a) and (b), respectively.

Focusing on the leading contribution to $\lambda_{L}$ in our large-$N$
expansion allows to perform a series of simplifications of Eq.~\ref{eq:bse}.
First, we set the fermionic propagators on mass-shell which requires
a representation in diagonal band-basis and which implicitly assumes
that a quasi-particles description is applicable. We therefore introduce
the projection operator ${\mathcal{P}}_{\alpha\beta}^{a}\left(\mathbf{k}\right)$ for the two bands $a=\pm 1$ which connects the pseudo-spin- and band-basis
by $G_{\alpha\beta}^{R}\left(k\right)=\sum_{a}\mathcal{P}_{\alpha\beta}^{a}\left(\mathbf{k}\right)G_{a}^{R}\left(k\right)$.
The projection operator has the properties
$\sum_{\alpha}\mathcal{P}^{a}_{\alpha\alpha}\left(\mathbf{k}\right)=1$
and $\sum_{\beta}\mathcal{P}_{\alpha\beta}^{a}\left(\mathbf{k}\right)\mathcal{P}_{\beta\gamma}^{b}\left(\mathbf{k}'\right)=\frac{1}{2}\mathcal{P}_{\alpha\gamma}^{a}\left(\mathbf{k}\right)(1+ab\frac{\mathbf{k}\cdot\mathbf{k}'}{|\mathbf{k}||\mathbf{k}'|})$, for $a,b=\pm 1$.
This allows us to replace the product of Green's functions by 
\begin{multline}
G_{\alpha\gamma}^{R}(\omega+\epsilon,\mathbf{k})G_{\delta\beta}^{A}\left(\epsilon,\mathbf{k}\right)\\
=2\pi i\frac{\mathcal{P}_{\alpha\gamma}^{a}(\mathbf{k})\mathcal{P}_{\delta\beta}^{b}(\mathbf{k})\delta\left(\epsilon-av|\mathbf{k}|\right)}{\omega-\left(a-b\right)|\mathbf{k}|+i0^{+}}.\label{eq:greens-product}
\end{multline}
To focus on the most rapidly growing term, we restrict Eq. \ref{eq:greens-product}
to $a=b$ and set the frequency of the scattering vertex in Eq. \ref{eq:scattering-vertex}
to zero, $\Gamma_{\gamma\delta}^{\alpha\beta}(\omega,k,k')|_{\omega=0}$.

Furthermore, to leading order in $1/N$, the squared anticommutator
is expressed by one band index only 
\begin{equation}
f_{\beta}^{\alpha}\left(\omega,k\right)=\sum_{a=\pm1}f_{a}\left(\omega,\mathbf{k}\right)\mathcal{P}_{\alpha\beta}^{a}\left(\mathbf{k}\right)2\pi\delta(\epsilon-av|\mathbf{k}|).\label{eq:f-w-k-2}
\end{equation}
Exploiting particle-hole symmetry, determining the Lyapunov exponent is eventually
reduced to solving the integral equation 
\begin{equation}
f(\omega,\mathbf{k})=\frac{i}{\omega}\frac{1}{N}\Big(1+\int_{\mathbf{k}'}M(\mathbf{k},\mathbf{k}')\, f(\omega,\mathbf{k}')\Big)\label{eq:integral-equation}
\end{equation}
where $f(\omega,\mathbf{k})=\int_{\epsilon}\sum_{\alpha}f_{\alpha}^{\alpha}(\omega,k)$
with the kernel $M=M_{+}+M_{-}$ comprised out of band-preserving
($+$) and band-changing ($-$) processes 
\begin{multline}
M_{\text{\ensuremath{\pm}}}(\mathbf{k},\mathbf{k}')=i\, K_{\pm+}\left(\mathbf{k},\mathbf{k}'\right)D_{ul}^{K}(|\mathbf{k}'|\mp|\mathbf{k}|,\mathbf{k}'\mp\mathbf{k})\\
+\sum_{a'b'}\int_{\tilde{k}}K_{+b'}(\mathbf{k}',\mathbf{k}'-\tilde{\mathbf{k}})\, K_{\pm a'}(\mathbf{k},\mathbf{k}-\tilde{\mathbf{k}})\\
\times G_{a'}^{ul}(\text{\ensuremath{\pm}}a'|\mathbf{k}|-\tilde{\epsilon},\mathbf{k}-\tilde{\mathbf{k}})G_{b'}^{lu}(b'|\mathbf{k}'|-\tilde{\epsilon},\mathbf{k}'-\tilde{\mathbf{k}})\\
\times D^{R}(\tilde{\epsilon},\tilde{\mathbf{k}})D^{A}(\tilde{\epsilon},\tilde{\mathbf{k}})\label{eq:M}
\end{multline}
where $K_{ab}\left(\mathbf{k},\mathbf{k}'\right)\equiv\frac{1}{2}(1+ab\frac{\mathbf{k}\cdot\mathbf{k}'}{|\mathbf{k}||\mathbf{k}'|})$.
The first term represents one-rung (denoted by $M_{\pm}^{(1)}$ in
the following) and the second term two-rung scattering, see Fig. \ref{fig:Diagrammatic-representation-of-f}.

The Lypanuov exponent is finally determined by finding the set of
eigenvalues $\{\lambda\}$ and corresponding eigenfunctions $\{f'\}$
of $M(\textbf{k},\textbf{k}')$. Being real and symmetric, it is possible to bring the kernel in diagonal form,
\begin{multline}
M'(\textbf{k},\textbf{k}')=\\
\int_{\textbf{q}\textbf{q}'}V(\textbf{k},\textbf{q})M(\textbf{q},\textbf{q}')V(\textbf{q}',\textbf{k}')=N\lambda_{\textbf{k}}\delta(\textbf{k}-\textbf{k}')\label{eq:M-decomposition}
\end{multline}
where the linear orthonormal transformation is described by the orthogonal
matrix $V$. Using Eq. \ref{eq:integral-equation}, we get 
\begin{equation}
f'(\textbf{k},\omega)=\frac{i}{\omega}\frac{1}{N}\left[e_{\textbf{k}}+N\lambda_{\textbf{k}}f'(\textbf{k},\omega)\right]
\end{equation}
where $f'(\textbf{k},\omega)=\int_{\textbf{q}}V(\textbf{k},\textbf{q})f(\textbf{q},\omega)$
and $e_{\textbf{k}}=\int_{\textbf{q}}V(\textbf{k},\textbf{q})$. Applying
the inverse Fourier transform $\int\frac{d\omega}{2\pi}\frac{e^{-i\omega t}}{(\omega+i0^{+})^{n+1}}=(-i)^{n+1}\theta(t)\frac{t^{n}}{n!}$,
we find 
\begin{equation}
f'(\textbf{k},t)=\frac{e_{\textbf{k}}}{N}\sum_{n=0}^{\infty}\frac{(\lambda_{\textbf{k}}t)^{n}}{n!}=\frac{e_{\textbf{k}}}{N}e^{\lambda_{\textbf{k}}t}
\end{equation}
which represents the desired exponential growth behavior described
by the spectrum of growth exponents $\lambda_{\textbf{k}}$ (see also Appendix \ref{LyapunovSpectr}).

The Lyapunov exponent is now defined as $\lambda_{L}=\max[\lambda_{\textbf{k}}]$,
and the corresponding eigenfunction is denoted $f'_{L}$. 
Consequently, the Lyapunov exponent can be efficiently determined as the largest eigenvalue of the eigenvalue equation 
\begin{equation}
\lambda\, f(\textbf{k},\omega)=\int_{\textbf{k}'}M(\mathbf{k},\mathbf{k}')f(\mathbf{k}',\omega)\label{eq:homo-bse}
\end{equation}
instead of solving the inhomogeneous Eq. \ref{eq:integral-equation}.
An explicit representation
of the homogeneous Bethe-Salpeter equation, which is solved numerically,
can be found in App. \ref{app:homo-bse}.
\begin{figure}[t]
\centering{}\includegraphics[width=1\columnwidth]{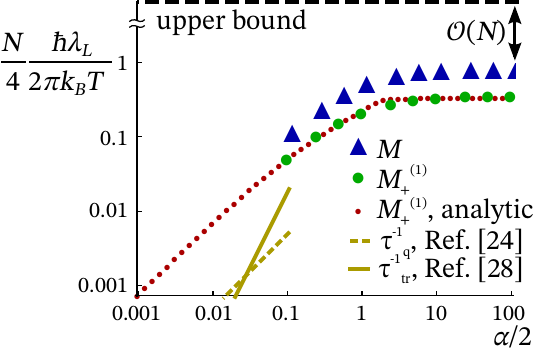}\protect\caption{\label{fig:Lyapunov-exponent}The Lyapunov exponent as a function
of coupling $\alpha$ to leading order in $1/N$: Blue triangles and
green dots represent values obtained using the full kernel $M$ and
focusing on the one-rung contribution $M_{+}^{(1)}$, respectively. We use a grid of $2^{9}\times2^{9}$
points in momentum space. These results are compared to values (red
dots) obtained by an analytic approximation for $M_{+}^{(1)}$ of
Eq. \ref{eq:BSE-with-Mplus} which is solved iteratively with much
higher precision. A linear function is fitted to the data points for
$\alpha\ll1$. We find that $\lambda_{L}=C\,\alpha\frac{4}{N}\frac{2\pi k_{B}T}{\hbar}$
with $C=0.37$ (see Appendix \ref{NumericalProcedure} for more details on the numerical procedure). The
solid and dashed yellowish lines represent the dephasing rate $\tau_{q}^{-1}$
evaluated at $\epsilon\approx\alpha k_{B}T$ and the transport rate
$\tau_{\text{tr}}^{-1}$ obtained from Ref. \cite{Schuett2011} and
Ref. \cite{Fritz2008}, respectively. }
\end{figure}

\subsection{Results}

The Lyapunov exponent as a function of coupling $\alpha$ is depicted
in Fig. \ref{fig:Lyapunov-exponent}. It saturates for strong coupling
to the asymptotic value given in Eq. \ref{eq:lambda-strong}. Even
if we extrapolate the number of fermion flavors to its physical value
$N=4$ the mentioned bound is not saturated. For weak coupling, we
obtain Eq. \ref{eq:lambda-weak}.

\begin{figure}[t]
\centering{}\includegraphics[width=1\columnwidth]{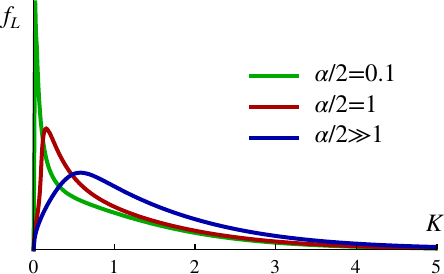}\protect\caption{\label{fig:Representative-eigenfunctions-co}Eigenfunctions corresponding
to the largest eigenvalue $\lambda_{L}$ as a function of energy for
three different couplings. In the plot, $K=\frac{v|\mathbf{k}|}{2k_{B}T}$.
In the case of strong coupling, the eigenfunction is peaked at energies
$v|\mathbf{k}|\approx k_{B}T$ which is the only energy scale present.
For weak coupling, the peak shifts to $v|\mathbf{k}|\propto\alpha k_{B}T$
which relates to an energy scale induced by finite coupling.}
\end{figure}

For our subsequent discussion it is important to determine the corresponding
eigenfunctions of the kernel $M(\mathbf{k},\mathbf{k}')$. We find
that in the case of strong coupling the eigenfunction $f_{L}\left(|\mathbf{k}|\right)$
is peaked at energies of order of the temperature, $v|\mathbf{k}|\approx k_{B}T$,
which is the only energy scale present, see Fig. \ref{fig:Representative-eigenfunctions-co}.
In the weak coupling regime however, the peak shifts due to the finiteness
of the coupling to $v|\mathbf{k}|\propto\alpha k_{B}T$ which is the
scale associated with the thermal screening of the Coulomb interaction
\cite{Schuett2011}.

As shown in Fig. \ref{fig:Lyapunov-exponent}, the dominant contribution
to scrambling in graphene is the one-rung band-preserving scrambling
process $M_{+}^{(1)}$. The Bethe-Salpeter equation only taking into
account $M_{+}^{(\text{1})}$ is given by 
\begin{alignat}{1}
\lambda f(\omega,K) & =\frac{2\pi k_{B}T}{\hbar}\frac{4}{N}\,\frac{2}{K}\int_{0}^{\infty}\frac{dK'}{2\pi}\int_{|K-K'|}^{|K+K'|}\frac{QdQ}{2\pi}\label{eq:BSE-with-Mplus}\\
\times & \frac{\sqrt{\left(K+K'\right)^{2}-Q^{2}}}{\sqrt{Q^{2}-\left(K-K'\right)^{2}}}\frac{\text{Im}\mathcal{D}^{R}\left(|K-K'|,Q\right)}{\sinh\left(|K-K'|\right)}\, f(\omega,K')\nonumber 
\end{alignat}
where we introduced the dimensionless momenta $K=\frac{v|\mathbf{k}|}{2k_{B}T}$
and the dimensionless imaginary part of the propagator $\text{Im}\mathcal{D}^{R}$
(see Appendix \ref{TheBSGlDetails} for its definition). In the co-scattering
limit $K\sim K'$, and transferred momenta smaller than the thermal
screening scale $Q\leq L_{\text{s }}^{-1}\ll1$ where $L_{\text{s }}^{-1}=l_{s}^{-1}\hbar v/k_{B}T=\alpha\,\ln2$,
the kernel (up to phase space factors) is 
\begin{multline}
\frac{2Q\,\text{Im}\mathcal{D}_{ul}^{R}\left(|K-K'|,Q\right)}{\sinh\left(|K-K'|\right)}\overset{K\rightarrow K'}{=}\frac{\alpha^{2}\ln2\,}{\left(Q+\alpha\ln2\right)^{2}}
\end{multline}
which approaches $1/\ln2$ for $Q\ll\alpha$. We note that the dressed
interaction becomes independent of coupling due to thermal screening
processes \cite{Schuett2011}. Thus, the dominating scrambling takes
place on orders of the screening scale $L_{\text{s}}^{-1}$ induced
by finite coupling. This is reflected in the overall result for the
Lyapunov exponent: The exponent is linear in coupling due to interactions
of order $\alpha^{0}$ which holds for transferred energies smaller
than the screening scale $L_{\text{s}}^{-1}\propto\alpha$. 

In order to interpret our result, Eq. \ref{eq:lambda-weak} for weak
coupling, we compare $\lambda_{L}$ with the characteristic energy
scales discussed in \cite{Schuett2011}. This discussion is most transparent
if we focus on Eq. \ref{eq:BSE-with-Mplus}. The kernel behaves qualitatively
as the one for the relaxation rate $\tau_{q}^{-1}\left(\epsilon\right)$
of \cite{Schuett2011}, i.e. there are no $1-\cos^{2}\theta_{\mathbf{k},\mathbf{k}'}$
back-scattering corrections that enter the transport rate $\tau_{\text{tr}}^{-1}\left(\epsilon\right)$
or energy weights $\propto(K-K')^{2}$ that determine the energy relaxation
rate $\tau_{E}^{-1}\left(\epsilon\right)$, respectively. For details
see also Fig. \ref{fig:Qualitative-representation-the} and the Appendix
\ref{sec:Characteristic-Rates}. Furthermore, our eigenfunctions vary
on a scale $\alpha k_{B}T$, see Fig. \ref{fig:Representative-eigenfunctions-co}.
Projection to energy scales $\alpha k_{B}T$ amounts essentially to
setting the typical energy scale $\epsilon\approx\alpha k_{B}T$.
In this limit follows indeed from Ref. \cite{Schuett2011} that $\tau_{q}^{-1}(\epsilon=\alpha k_{B}T)\approx0.58\alpha k_{B}T/\hbar N$
similar to our scrambling rate. While there are differences in the
detailed numerical prefactors -- the coefficient of $\lambda_{L}$
is about 16 times larger, see Fig. \ref{fig:Lyapunov-exponent} --
the scrambling rate in graphene behaves as a dephasing rate. For weak
coupling this scale is much faster than the characteristic transport
collision rate of the hydrodynamic regime $\tau_{{\rm tr}}^{-1}\sim\alpha^{2}k_{B}T/\hbar N$,
guaranteeing local thermalization which is a key prerequisite of a
hydrodynamic description. Since $\lambda_{L}\gg\tau_{E}^{-1}$ for
energies between $\alpha k_{B}T$ and $k_{B}T$ we also find that
actual thermalization is a much slower processes than information
scrambling.

\section{Kinetic equation\label{sec:Kinetic-equation}}

In this section we present an alternative approach to describe the
spread of information in time and space in many-body systems.
In contrast to the diagrammatic expansion conducted in the previous
section, we show that scrambling is described by an integro-differential
equation similar to the well-known Boltzmann equation. It describes
the growth of an initially small, localized perturbation. 

The process of information scrambling is governed by two scales: The
Lyapunov exponent $\lambda_{L}$ and the Butterfly velocity $v_{B}=\sqrt{4D\lambda_{L}}$
which gives rise to an additional length scale $l_{B}=\frac{v_{B}}{\lambda_{L}}$
associated with the spatial spreading of information. 
For the system discussed in this work, the scrambling parameters are small
$\lambda_{L},l_{B}^{-1}\sim\mathcal{O}\left(1/N\right)$. Based on
the smallness of these parameters we propose that the spreading of
information and the exponential growth signaling chaotic behavior
is described by a quantity $f(t,\mathbf{x})$ which is governed by
the partial differential equation 
\begin{equation}
\partial_{t}f-D\Delta_{\mathbf{x}}f=f_{0}\delta\left(t\right)\delta\left(\mathbf{x}\right)+\lambda_{L}f,\label{eq:pde-f}
\end{equation}
where higher order gradient terms are suppressed in higher orders
of $\lambda_{L}$ and $l_{B}^{-1}$. The LHS represents a diffusion
equation characterized by diffusion constant $D$, whereas the RHS
contains a source term and a term $\sim\lambda_{L}$ that indicates
that $f$ is not a conserved quantity: A perturbing term $f_{0}\ll1$
triggering the onset of growth and a second term causing the characteristic
exponential growth behavior. Eq. \ref{eq:pde-f} is valid for early
times, i.e. $0<t-\frac{|\mathbf{x}|}{v_{B}}<t^{*}$ with the scrambling
time $t^{*}\sim\lambda^{-1}\log N$. For times $t-\frac{|\mathbf{x}|}{v_{B}}>t^{*},$
non-linear terms are relevant causing $f$ to saturate against its
asymptotic long-time value. 

The solution of Eq. \ref{eq:pde-f} is obtained by Fourier transform
and is given by 
\begin{equation}
f(\mathbf{x},t)=\frac{f_{0}}{4\pi Dt}e^{\lambda_{L}t-\frac{\mathbf{x}^{2}}{4 Dt}}\approx\frac{f_{0}}{4\pi Dt}e^{2\lambda_{L}\left(t-\frac{|\mathbf{x}|}{v_{B}}\right)}\label{eq:f-sol}
\end{equation}
where the approximative result is obtained for $|\mathbf{x}|\approx v_{B}t$.
It suggests that information spreads diffusively. However, due to
the additional source term in Eq. \ref{eq:pde-f} spreading is enhanced
and the perturbation propagates 'quasi-ballistically' as indicated
by the approximative solution Eq. \ref{eq:f-sol}. 

In the following section we derive Eq. \ref{eq:pde-f} microscopically
for the specific case of graphene. The approach is however more general
and also applicable to other systems which can be treated perturbatively.

\subsection{Derivation}

We start by introducing a generating field $W$ which allows us to
express the correlation function containing OTOCs introduced in Eq.
\ref{eq:f(x)} as functional derivative. It is introduced by directly
coupling to the inter-loop fermionic density 
\begin{equation}
S'[W]=-\frac{1}{N}\int_{x}\sum_{i\alpha}W_{i\alpha}(x_{1},x_{2})\bar{\psi}_{i\alpha}^{l\text{cl}}(x_{1})\psi_{i\alpha}^{u\text{q}}(x_{2})
\end{equation}
which enters as a contribution to the effective two-loop Keldysh action
$S'_{\mathcal{K}}=S_{\mathcal{K}}+S'$. Consequently, the correlation
function is obtained as a functional derivative of the fermionic inter-loop
Keldysh component $G_{lu,\alpha\beta}^{K}(x_{1},x_{2};W)=-i\langle\psi_{i\alpha}^{l\text{cl}}(x_{1})\bar{\psi}_{i\beta}^{u\text{q}}(x_{2})\rangle$:
\begin{equation}
f_{\beta\gamma}^{\alpha\gamma}(x_{1},x_{2},x_{3},0)=-\frac{1}{N}\frac{\delta G_{lu,\alpha\beta}^{K}(x_{1},x_{3};W)}{\delta W_{j\gamma}(x_{2},0)}\bigg|_{W=0}\label{eq:var-G}
\end{equation}
We are therefore interested in the dependence of $G_{lu}^{K}$ on
the field $W$, which can be interpreted as the evolution of $G_{lu}^{K}$
in time and spatial space due to an earlier localized perturbation.

\begin{figure}[t]
\centering{}\centering{}\includegraphics[width=1\columnwidth]{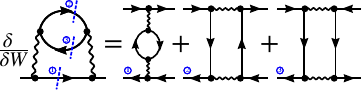}\protect\caption{\label{fig:self-energy-diagram}Leading order self-energy diagram
in the large-$N$, weak-coupling ($\alpha\ll1$) regime. Dashed blue
lines represent functional derivatives applied to obtain the linearized
kinetic equations in Eq. \ref{eq:final-f-pde}. }
\end{figure}

The dynamics of the Keldysh components are conveniently described
by means of kinetic equations, an established approach to non-equilibrium
problems (see e.g. Ref. \cite{Kamenev2011}). We therefore introduce
the usual parametrization of $G^{K}$ applied to the inter-loop Keldysh
components as\begin{subequations} 
\begin{eqnarray}
G_{lu,a}^{K} & = & G_{a}^{R}\circ H_{a}^{lu}-H_{a}^{lu}\circ G_{a}^{A},\\
G_{ul,a}^{K} & = & G_{a}^{R}\circ H_{a}^{ul}-H_{a}^{ul}\circ G_{a}^{A},
\end{eqnarray}
\end{subequations}where conceptually non-equilibrium aspects are
stored in the inter-loop distribution functions $H_{a}^{lu}\left(x_{1},x_{2}\right)$
and $H_{a}^{ul}\left(x_{1},x_{2}\right)$, $a$ represents the band-index
and $\circ$ stands for a convolution with regards to time and spatial
space. In equilibrium, $H^{lu,\left(\text{eq}\right)}\left(\epsilon,\mathbf{k}\right)=\cosh^{-1}(\frac{\epsilon}{2k_{B}T})=-H^{ul,\left(\text{eq}\right)}\left(\epsilon,\mathbf{k}\right)$,
as already indicated in Eq. \ref{eq:fermionic-inter-loop-components}.
Within the concept of single particle self-energies, equations of
motions of inter-loop distributions functions are obtained by using
Dyson's equation for intra- and inter-loop single-particle propagator
components and are given in the presence of the generating field by\begin{subequations}\label{eq:qbe}
\begin{alignat}{1}
-\left[G_{a}^{-1}\overset{\circ}{,}H_{a}^{lu}\right] & =\notag\\
\frac{W}{N}+ & \Sigma_{lu,a}^{K}-\left(\Sigma_{a}^{R}\circ H_{a}^{lu}-H_{a}^{lu}\circ\Sigma_{a}^{A}\right),\label{eq:qbe1}\\
-\left[G_{a}^{-1}\overset{\circ}{,}H_{a}^{ul}\right] & =\Sigma_{ul,a}^{K}-\left(\Sigma_{a}^{R}\circ H_{a}^{ul}-H_{a}^{ul}\circ\Sigma_{a}^{A}\right).\label{eq:qbe2}
\end{alignat}
\end{subequations}The expression for the inter-loop Keldysh self-energy
is given by $\Sigma_{lu}^{K}=\frac{i}{2N}G_{lu}^{K}\left(x_{1},x_{2}\right)D_{ul}^{K}\left(x_{1},x_{2}\right)$,
whereas the retarded and advanced components are as indicated in  Sec.
\ref{sec:The-Model}. For convenience, we replace the bosonic Keldysh
components by $D_{ul}^{K}=-D^{R}\circ\Pi_{ul}^{K}\circ D^{A}$ with
the inter-loop Keldysh polarization operator $\Pi_{ul}^{K}\left(x_{1},x_{2}\right)=\frac{i}{2}G_{ul}^{K}\left(x_{1},x_{2}\right)G_{lu}^{K}\left(x_{2},x_{1}\right)$
to eliminate the bosonic Keldysh components. This expansion holds
in the large-$N$, weak coupling limit, which shows the interesting
behavior as discussed in the previous section. A diagrammatic representation
of the considered self-energy contribution is depicted in Fig. \ref{fig:self-energy-diagram}.

In the derivation of the equations of motions Eq. \ref{eq:qbe} it
was furthermore implicitly assumed that in the presence of interactions
the band index remains a 'good' quantum number, i.e.~the unitary transformation
represented by the matrix $\left(\mathcal{U}_{\alpha a,\textbf{k}}\right)$,
which diagonalizes the free single-particle $\sum_{a}\mathcal{U}_{\alpha a,\mathbf{k}}^{\dagger}G_{a}(k)\mathcal{U}_{a\beta,\mathbf{k}}=G_{\alpha\beta}(k)$
model, diagonalizes also the self-energy expressions $\sum_{a}\mathcal{U}_{\alpha a,\mathbf{k}}^{\dagger}\Sigma_{a}(k)\mathcal{U}_{a\beta,\mathbf{k}}=\Sigma_{\alpha\beta}(k)$.
This assumption proves legitimate in the low-temperature limit \cite{Schuett2011}. 

As a next step, we introduce center-of-mass coordinates $X=\frac{1}{2}(x_{1}+x_{2})$
and derivations thereof $x=x_{1}-x_{2}$, respectively, and replace
all quantities by quantities depending on these coordinates $Q(x_{1},x_{2})\rightarrow\tilde{Q}\left(X,x\right)$.
The Wigner-transform is subsequently introduced as $\tilde{Q}(X,p)=\int_{x}e^{i(\omega t-\textbf{p}\cdot\textbf{x})}\tilde{Q}(X,x)$.
Gradient terms, which are generated by Wigner-transforming convolution
terms, are assumed to be small as argued in the introductory part of
this section resulting in \begin{subequations}\label{eq:q-boltz}
\begin{alignat}{1}
-iZ^{-1}\left(\partial_{T}+\mathbf{v}_{\mathbf{p}}^{*}\cdot\nabla_{\mathbf{X}}\right)\tilde{H}_{a}^{lu} & =\nonumber \\
\frac{\tilde{W}}{N}+\tilde{\Sigma}_{lu,a}^{K}-\tilde{H}_{a}^{lu}\big(\tilde{\Sigma}_{a}^{R} & -\tilde{\Sigma}_{a}^{A}\big)+F[\Delta\tilde{H}],\\
-iZ^{-1}\left(\partial_{T}+\mathbf{v}_{\mathbf{p}}^{*}\cdot\nabla_{\mathbf{X}}\right)\tilde{H}_{a}^{ul} & =\nonumber \\
\tilde{\Sigma}_{ul,a}^{K}-\tilde{H}_{a}^{ul}\big(\tilde{\Sigma}_{a}^{R} & -\tilde{\Sigma}_{a}^{A}\big)+F[\Delta\tilde{H}],
\end{alignat}
\end{subequations}where $\tilde{H}_{a}(T,\mathbf{X},\epsilon,\mathbf{p})$
denote the Wigner-transformed distribution functions which are assumed,
besides the external generating field $\tilde{W}$, to be the only
quantities depending on center-of-mass coordinates. First order gradient
terms renormalize the single-particle parameters, such as the quasi-particle
weight $Z=(1-\partial_{\epsilon}\text{Re}\tilde{\Sigma}_{a}^{R})^{-1}$
($Z\approx1$ in the case of graphene \cite{Schuett2011}) and the
renormalized group velocity $\mathbf{v}_{\mathbf{p}}^{*}=v_{F}Z\nabla_{\mathbf{p}}(\epsilon_{\mathbf{p}}+\text{Re}\tilde{\Sigma}_{a}^{R})$,
whereas terms of second order gradients are stored in the term $F[\Delta_{\mathbf{x}}\tilde{H}]$.
These second order gradient terms eventually give rise to spatial
gradient terms as postulated in Eq. \ref{eq:pde-f}, which are however
dropped in the ongoing discussion as their explicit form does not
yield any new insights. 

The single-particle spectral function $A_{a}\left(\epsilon,\mathbf{p}\right)=-\text{Im}\tilde{G}_{a}^{R}\left(\epsilon,\mathbf{p}\right)$
is peaked for $\epsilon\approx\epsilon_{\mathbf{p}}$. If the momentum
dependence of $\text{Im}\Sigma_{a}^{R}$ is negligible, which is the
case for graphene \cite{Schuett2011}, the spectrum contains no incoherent
background and the quasi-particle description applies \cite{Woelfle2018}.
This allows us to integrate out the frequency dependence to define
the quasi-particle distribution function 
\begin{eqnarray}
h_{a\textbf{p}}(T,\mathbf{X}) & = & \int_{\epsilon}2A_{a}\left(\epsilon,\mathbf{p}\right)\tilde{H}_{a}\left(T,\mathbf{X},\epsilon,\mathbf{p}\right).\label{eq:hap}
\end{eqnarray}
In the following we, approximate $A_{a}\left(\epsilon,\mathbf{p}\right)\approx\pi\delta\left(\epsilon-\epsilon_{\mathbf{p}}\right)$
and conduct the frequency integration which is identical to the mass-shell
approximation conducted in the previous section withing the diagrammatic
approach, see Eq. \ref{eq:greens-product} and \ref{eq:f-w-k-2}. 

We eventually arrive at the following set of coupled partial differential
equations\begin{subequations}\label{eq:bmann-eq} 
\begin{alignat}{1}
(\partial_{T}\negthinspace+\negthinspace\mathbf{v}_{\mathbf{p}}^{*}\cdot\nabla_{\mathbf{X}})h_{a\mathbf{p}}^{lu}\left(T,\mathbf{X}\right)\negthinspace-\negthinspace\frac{W_{a\mathbf{p}}}{N} & =I\left[h^{lu},h^{ul}\right],\\
\left(\partial_{T}+\mathbf{v}_{\mathbf{p}}^{*}\cdot\nabla_{\mathbf{X}}\right)h_{a\mathbf{p}}^{ul}\left(T,\mathbf{X}\right) & =I\left[h^{ul},h^{lu}\right],
\end{alignat}
\end{subequations}where the mass-shell restricted generating field
reads $W_{a\mathbf{p}}\left(T,\mathbf{X}\right)=i\int_{\epsilon}2A_{a}\left(\epsilon,\mathbf{p}\right)W\left(T,\mathbf{X},\epsilon,\mathbf{p}\right)$.
In analogy to the quasi-classical Boltzmann equation, we introduced
a 'collision-integral' $I[h^{lu},h^{ul}]$ (and conversely $I[h^{ul},h^{lu}]$
which is obtained by exchanging $h^{ul}\leftrightarrow h^{lu}$) on
the RHSs of the previous equation which represents interaction processes
coupling the different components of $h^{ul}$ and $h^{lu}$. It is
given by 
\begin{multline}
I\left[h^{lu},h^{ul}\right]=\\
\negthinspace-\frac{\pi}{2N}\negthinspace\negthinspace\sum_{a'bb'}\int_{\mathbf{qk}}\negthinspace\negthinspace\negthinspace\negthinspace\delta(\epsilon_{a\mathbf{p}}\negthinspace-\negthinspace\epsilon_{b\mathbf{k}}\negthinspace+\negthinspace\epsilon_{b'\mathbf{q+k}}\negthinspace-\negthinspace\epsilon_{a'\mathbf{p+q}})\negthinspace\negthinspace\times\negthinspace\negthinspace|T_{a'bab'}(\mathbf{p},\mathbf{k},\mathbf{q})|^{2}\\
\negthinspace\times\negthinspace\bigg[h_{a'\mathbf{p+q}}^{lu}h_{b'\mathbf{q+k}}^{ul}h_{b\mathbf{k}}^{lu}\negthinspace-\negthinspace\frac{h_{a'\mathbf{p+q}}^{lu,(\text{eq})}h_{b'\mathbf{q+k}}^{ul,(\text{eq})}h_{b\mathbf{k}}^{lu,(\text{eq})}}{h_{a\mathbf{p}}^{lu,(\text{eq})}}h_{a\mathbf{p}}^{lu}\bigg]
\end{multline}
where the $\delta$-function ensures energy conservation, the interaction
matrix-element $T$, which is given by $T_{a'bab'}(\mathbf{p},\mathbf{k},\mathbf{q})=\tilde{D}_{\mathbf{q}}(U_{\mathbf{p}}U_{\mathbf{p}+\mathbf{q}}^{\dagger})_{aa'}(U_{\mathbf{q}+\mathbf{k}}U_{\mathbf{k}}^{\dagger})_{b'b}$
in the case of graphene, and the equilibrium distribution functions
$h_{a\mathbf{p}}^{lu,(\text{eq})}=\cosh^{-1}(\frac{\epsilon_{a\mathbf{p}}}{2k_{B}T})=-h_{a\mathbf{p}}^{ul,(\text{eq})}$.
To derive the second term the intra-loop distribution functions were replaced by their equilibrium value as being not affected by $W_{a\textbf{p}}$ and it was used that
\begin{multline} 
h^{(\text{eq})}_{b'\mathbf{q+k}}h^{(\text{eq})}_{b\mathbf{k}}-h^{(\text{eq})}_{a'\mathbf{p+q}}h^{(\text{eq})}_{b\mathbf{k}}+h^{(\text{eq})}_{a'\mathbf{p+q}}h^{(\text{eq})}_{b'\mathbf{q+k}}-1 \\ = \frac{h_{a'\mathbf{p+q}}^{lu,(\text{eq})}h_{b'\mathbf{q+k}}^{ul,(\text{eq})}h_{b\mathbf{k}}^{lu,(\text{eq})}}{h_{a\mathbf{p}}^{lu,(\text{eq})} } 
\end{multline}
with the equilibrium intra-loop distribution function $h^{(\text{eq})}_{a\mathbf{k}} = \tanh(\frac{\epsilon_{a\mathbf{k}}}{2k_B T})  $. 
In equilibrium, the collision-integral vanishes, i.e. $I[h_{a\mathbf{p}}^{lu,(\text{eq})},h_{a\mathbf{p}}^{ul,(\text{eq})}]=I[h_{a\mathbf{p}}^{ul,(\text{eq})},h_{a\mathbf{p}}^{lu,(\text{eq})}]=0$,
which is the only solution for vanishing external field $W_{a\mathbf{p}}=0$.
Here, the second term, which is traced back to intra-loop self-energy
contributions (see Eq. \ref{eq:q-boltz}), is of particular significance.
It does not directly contribute to the process of scrambling, but
it is necessary to establish equilibrium. In contrast to the collision-integrals
of the conventional Boltzmann equation, the products of distribution
functions differ which results, e.g., in the absence of
an equivalent of the H-theorem, or conservation laws \cite{Aleiner2016}. 

To obtain the final result, we perform the functional derivative and
introduce  
\begin{eqnarray}
f_{a}\left(X,\mathbf{p}\right) & = & \sum_{b}\int_{\mathbf{k}}\frac{\delta h_{a\mathbf{p}}\left(X\right)}{\delta W_{b\mathbf{k}}\left(0\right)}\bigg|_{W=0},
\end{eqnarray}
which relates to the quantity $f_{a}\left(\omega,\mathbf{k}\right)$
found in Eq. \ref{eq:f-w-k-2} of the previous section. It relates
to the initial correlation Eq. \ref{eq:var-G} where one half of the
indices is traced over, and external legs are restricted to the same
band, which contributes predominantly to the exponential growth behavior
as shown in the previous section. Applied to the kinetic equations
we obtain\begin{subequations}\label{eq:final-f-pde}
\begin{alignat}{1}
(\partial_{T}\negthinspace+\negthinspace\mathbf{v}_{\mathbf{p}}^{*}\negthinspace\cdot\negthinspace\nabla_{\mathbf{X}})f_{a\mathbf{p}}^{lu}\left(T,\mathbf{X}\right)\negthinspace-\negthinspace\frac{\delta\negthinspace\left(T\right)\delta\negthinspace\left(\mathbf{X}\right)}{N} & \negthinspace=\negthinspace I[f^{lu},f^{ul}],\\
\left(\partial_{T}+\mathbf{v}_{\mathbf{p}}^{*}\cdot\nabla_{\mathbf{X}}\right)f_{a\mathbf{p}}^{ul}\left(T,\mathbf{X}\right)= & \negthinspace-\negthinspace I[f^{ul},f^{lu}],
\end{alignat}
\end{subequations}where the linearized collision-integral is given
by
\begin{multline}
I\left[f^{lu},f^{ul}\right]=\\
\frac{\pi}{2N}\negthinspace\negthinspace\sum_{a'bb'}\int_{\mathbf{qk}}\negthinspace\negthinspace\negthinspace\negthinspace\delta(\epsilon_{a\mathbf{p}}\negthinspace-\negthinspace\epsilon_{b\mathbf{k}}\negthinspace+\negthinspace\epsilon_{b'\mathbf{q+k}}\negthinspace-\negthinspace\epsilon_{a'\mathbf{p+q}})\negthinspace\negthinspace\times\negthinspace\negthinspace|T_{a'bab'}(\mathbf{p},\mathbf{k},\mathbf{q})|^{2}\\
\negthinspace\times\negthinspace\Big[h_{b'\mathbf{q+k}}^{\left(0\right)}h_{b\mathbf{k}}^{\left(0\right)}\, f_{a'\mathbf{p+q}}^{lu}\negthinspace+\negthinspace h_{a'\mathbf{p+q}}^{\left(0\right)}h_{b'\mathbf{q+k}}^{\left(0\right)}\, f_{b\mathbf{k}}^{lu}\\
-\negthinspace h_{a'\mathbf{p+q}}^{\left(0\right)}h_{b\mathbf{k}}^{\left(0\right)}\, f_{b'\mathbf{q+k}}^{ul}\negthinspace-\negthinspace\frac{h_{a'\mathbf{p+q}}^{\left(0\right)}h_{b'\mathbf{q+k}}^{\left(0\right)}h_{b\mathbf{k}}^{\left(0\right)}}{h_{a\mathbf{p}}^{\left(0\right)}}\, f_{a\mathbf{p}}^{lu}\Big],
\end{multline}
 and $h_{a\mathbf{p}}^{\left(0\right)}=h_{a\mathbf{p}}^{lu,(\text{eq})}=-h_{a\mathbf{p}}^{ul,(\text{eq})}$.
The first three terms entering the collision integral are diagrammatically
represented in Fig. \ref{fig:self-energy-diagram} where each contribution
is obtained by performing the functional derivative, i.e. by 'cutting'
one solid line, respectively. In comparison to our diagrammatic approach,
the first term represents the one-rung contribution, whereas the second
and third term the two-rung contribution. The last term of the collision
term is due to intra-loop contributions and does not contribute to
scrambling. 

\subsection{Connection to the diagrammatic approach}
To make the connection to the diagrammatic approach more explicit,
we show the identity of the first term in the collision integral with
the one-rung diagrammatic contribution as discussed in Eq. \ref{eq:integral-equation}.
By performing the Fourier transform and dropping the spatial gradient
term as well as all terms except the first term of the collision integral,
we find by using $K_{ab}(\mathbf{p},\mathbf{k})=(U_{\mathbf{p}}U_{\mathbf{k}}^{\dagger})_{ab}(U_{\mathbf{k}}U_{\mathbf{p}}^{\dagger})_{ba}$:
\begin{multline}
-i\omega f_{a\mathbf{p}}^{lu}(\omega)=\frac{1}{N}\Big(1+\negthinspace\sum_{a'}\negthinspace\int_{\mathbf{q}}\negthinspace\negthinspace K_{aa'}\negthinspace\left(\mathbf{p},\mathbf{p+q}\right)\\
\times\tilde{D}_{ul}^{K}\left(\negthinspace\mathbf{q},\epsilon_{a'\mathbf{p+q}}\negthinspace-\negthinspace\epsilon_{a\mathbf{p}}\right)\negthinspace f_{a'\mathbf{p+q}}^{lu}(\omega)\Big)
\end{multline}
where we introduced
\begin{multline}
\tilde{D}_{ul}^{K}\left(\mathbf{q},\omega\right)=\frac{\pi}{2}|D\left(\mathbf{q}\right)|^{2}\sum_{bb'}\int_{\mathbf{k}}\delta\left(\epsilon_{b'\mathbf{q+k}}-\epsilon_{b\mathbf{k}}-\omega\right)\\
K_{bb'}\left(\mathbf{k},\mathbf{k+q}\right)h_{b'\mathbf{q+k}}^{\left(0\right)}h_{b\mathbf{k}}^{\left(0\right)}.
\end{multline}
This is identical to Eq. \ref{eq:integral-equation} obtained in
the diagrammatic approach. 

To connect Eq. \ref{eq:final-f-pde} to \ref{eq:pde-f}, one has to
solve for $f^{ul}$ and $f^{lu}$ which is, as outlined in the previous
section, a demanding task. The solution consequently gives expressions
for the parameters $\lambda_{L}$ and $D$ and the corresponding eigenfunction
$f$ describing the process of information scrambling. Note that the
order one gradient terms on the LHSs of Eq. \ref{eq:final-f-pde}
vanish when taking the average of the external momentum angle: By
assuming $f_{\mathbf{p}}=f_{|\mathbf{p}|}$, which is legitimate for
a rotational symmetric initial perturbation, and $\mathbf{v}_{\mathbf{p}}\propto\frac{\mathbf{p}}{|\mathbf{p}|}$,
averaging yields $\int_{0}^{2\pi}\frac{d\phi}{2\pi}\thinspace\mathbf{v}_{\mathbf{p}}\cdot\nabla f_{\mathbf{p}}\left(T,\mathbf{X}\right)=0$,
and Eq. \ref{eq:pde-f} is recovered.


In this section, we reproduced the results obtained earlier in a diagrammatic
approach using non-equilibrium techniques. This puts the previously
obtained results on firmer ground, but also gives a deeper insight
into the theoretical description of information scrambling in many-body
systems: The free term entering the Bethe-Salpether equation can be
interpreted as perturbing source term of Eq. \ref{eq:pde-f}. This
quantum kinetic approach can be applied to other weak coupling problems
as well. 

We also comment on the experimental accessibility of scrambling. As shown in the kinetic equation-based formulation, the inter-loop distribution functions $h^{ul}$ and $h^{lu}$ are sensitive to the processes of scrambling. Their evolution is determined by Eq. \ref{eq:bmann-eq} where $h^{ul}$ and $h^{lu}$ as well as the intra-loop distribution function $h$ are coupled in the collision integral. The evolution of $h$ however is determined by a conventional kinetic equation (see e.g. Ref. \cite{Kamenev2011}) and is therefore not affected by the inter-loop distribution functions. Measuring $h^{ul}$ and $h^{lu}$ by measuring $h$ via physical observables is therefore not possible.

\section{Conclusion}

In this work, we determined the information scrambling rate $\lambda_{L}$
of graphene as a function of the coupling constant $\alpha$ of the
Coulomb interaction within a large-$N$ expansion. We showed that
for strong coupling ($\alpha\gg1$), scrambling saturates and is solely
determined by temperature which is the only energy scale present. 

In contrast at weak coupling ($\alpha\ll1$), new scales, such as
the thermal screening length $l_{s}^{-1}$ as discussed previously,
emerge. These additional scales cause physical quantities to scale differently with temperature
and coupling constant rendering them distinguishable. In this regime,
the scrambling rate scales as $\lambda_{L}\sim\alpha k_{B}T/\hbar N$
and is consequently much larger than the transport rate $\tau_{{\rm tr}}^{-1}\sim\alpha^{2}k_{B}T/\hbar N$
that occurs in the d.c.~conductivity or the electron viscosity. Instead,
the scrambling rate in graphene is closely related to the quantum
or dephasing scattering rate $\tau_{q}^{-1}\left(\epsilon,T\right)$
for characteristic energies $\epsilon\sim\alpha k_{B}T$ . Scrambling
processes at weak coupling are therefore a lot faster than the collisions
that give rise to the hydrodynamic behavior of graphene implying that
graphene is a comparatively fast information scrambler which is characterized
by single particle decay. However, $\lambda_{L}$, as defined by Eq.
\ref{eq:f(x)}, is not relevant for local thermalization as scrambling
probes the system only on a rather small range of excitation energies
around $\epsilon\sim\alpha k_{B}T$ and is parametrically larger than
the energy relaxation rate $\tau_{E}^{-1}\sim\alpha^{2}\log(\alpha^{-1})k_{B}T/\hbar N$
that one would expect to govern thermalization.

In the second part of this work we presented an alternative approach
towards scrambling in many-body systems. We showed that the results
obtained in a diagrammatic approach are reproduced by using non-equilibrium
techniques yielding a partial differential equation describing the
growth and spread of an initially small, localized perturbation. This
approach allows a physically deeper insight into the process of information
scrambling in many-body systems. 

\textit{Acknowledgements:} We thank R. Davison, I. V. Gornyi, Y. Gu,
A. D. Mirlin, P. M. Ostrovsky and S. Syzranov for fruitful discussions.
MS acknowledges support from the German National Academy of Sciences
Leopoldina through grant LPDS 2016-12.

\onecolumngrid

\appendix

\section{Scattering rates of electrons in graphene\label{sec:Characteristic-Rates}}

In the strong coupling limit ($\alpha\gg1$) all characteristic single
particle scattering rates are of order $k_{{\rm B}}T/\hbar N$. In
contrast in the weak coupling limit ($\alpha\ll1$), electrons in
graphene display a number of distinct scattering rates, such as the
dephasing or quantum rate $\tau_{q}^{-1},$ the rate for energy relaxations
$\tau_{E}^{-1}$ , or the rate relevant for transport processes $\tau_{{\rm tr}}^{-1}$,
which scale differently with coupling constant and temperature. The
origin of these scaling behaviors is the infrared-singular collision
kernel, a behavior to some extend similar to that of weakly interacting
diffusive electrons discussed in Ref. \cite{Patel2017_2}. A detailed
analysis of these frequency and energy-dependent scattering rates
was performed in Ref. \cite{Schuett2011}. The obtained results, which
are relevant for this work, are reviewed shortly in the following. 

The scattering rates are given by 
\begin{eqnarray}
\tau_{i}^{-1}\left(\epsilon,T\right) & = & \pi\int_{\omega}\left(\coth\left(\frac{\omega}{2k_{B}T}\right)+f\left(\epsilon-\omega\right)\right)\int_{\mathbf{q}}{\rm Im}D^R\left(\omega,\mathbf{q}\right)\nonumber \\
 & \times & {\cal K}_{i}\left(\mathbf{p\cdot}\mathbf{q},\omega\right)\sum_{s=\pm}\delta\left(\epsilon-\omega-v\left|\mathbf{p}-\mathbf{q}\right|\right).
\end{eqnarray}
where the specific rates distinguish themselves through different
kernels ${\cal K}_{i=\left\{ q,E,{\rm tr}\right\} }\left(\mathbf{p\cdot}\mathbf{q},\omega\right)$.
Here, $\mathbf{p}$ is the external momentum that enters the analysis
through $\epsilon=v\mathbf{\left|p\right|}$ on the mass shell. The
kernel for the dephasing rate is ${\cal K}_{q}=1$ and yields the
single particle scattering rate $\tau_{q}^{-1}\left(\epsilon,T\right)=-2{\rm Im}\Sigma^R\left(\epsilon,\mathbf{p}\right)$
with single particle self-energy $\Sigma^R\left(\epsilon,\mathbf{p}\right)$.
The energy relaxation rate is determined by ${\cal K}_{E}=\left(\omega/k_{B}T\right)^{2}$.
It is the relevant rate to determine the energy diffusion coefficient.
Finally, the transport scattering rate that enters transport coefficients
such as the electrical conductivity or the shear viscosity is determined
by ${\cal K}_{{\rm tr}}=\sin^{2}\theta_{\mathbf{p},\mathbf{q}}$ where
$\theta_{\mathbf{p},\mathbf{q}}$ is the angle between the momenta
$\mathbf{p}$ and $\mathbf{q}$. 

The main results of Ref. \cite{Schuett2011} are summarized as follows:
First, it is necessary to carefully distinguish between the relevant
energy regimes. For the dephasing rate holds that 
\begin{equation}
\tau_{q}^{-1}\left(\epsilon,T\right)\sim\left\{ \begin{array}{ccc}
\frac{1}{N}T\sqrt{\frac{\epsilon}{T}} & {\rm if} & \epsilon\ll\alpha^{2}T\\
\frac{\alpha}{N}T & {\rm if} & \epsilon\gg\alpha^{2}T
\end{array}\right.
\end{equation}
where we suppressed numerical coefficients of order unity, and $\hbar=k_{B}=1$
for the sake of representation. As shown in detail in Ref. \cite{Schuett2011},
the numerical coefficient in front of $\frac{\alpha}{N}T$ for $\epsilon \gg \alpha^2 T$ depends on whether $\epsilon$
is smaller or larger than the scale $\alpha T$, which for small $\alpha$
is large compared to $\alpha^{2}T$. This behavior is owed to the
screening length $l_{s}^{-1}\sim\alpha T/v$ due to thermally excited
carriers. 

The situation is significantly richer for the transport rate
\begin{equation}
\tau_{{\rm tr}}^{-1}\left(\epsilon,T\right)\sim\left\{ \begin{array}{ccc}
\frac{1}{N}T\sqrt{\frac{\epsilon}{T}} & {\rm if} & \epsilon\ll\alpha^{2}T\\
\frac{\alpha}{N}T & {\rm if} & \alpha^{2}T\ll\epsilon\ll\alpha T\\
\frac{\alpha^{2}}{N}T\left(\frac{T}{\epsilon}\right) & {\rm if} & \alpha T\ll\epsilon\ll T\\
\frac{\alpha^{2}}{N}T\left(\frac{T}{\epsilon}\right)^{2} & {\rm if} & T\ll\epsilon
\end{array}\right..
\end{equation}
If one uses this scattering rate as relevant input in the collision
integral of a kinetic equation, it holds that transport coefficients
are governed by the rate for the typical energies $\epsilon\sim T$,
where $\tau_{{\rm tr}}^{-1}\sim\frac{\alpha^{2}}{N}T$. 

Finally, for the energy relaxation rate holds that 
\begin{equation}
\tau_{E}^{-1}\left(\epsilon,T\right)\sim\left\{ \begin{array}{ccc}
\frac{1}{N}T\sqrt{\frac{\epsilon}{T}} & {\rm if} & \epsilon\ll\alpha^{2}T\\
\frac{\alpha^{2}}{N}T\sqrt{\frac{T}{\epsilon}}\log\left(\frac{\epsilon}{\alpha^{2}T}\right) & {\rm if} & \alpha^{2}T\ll\epsilon\ll T\\
\frac{\alpha^{2}}{N}T\left(\frac{\epsilon}{T}\right)^{\frac{3}{2}}\log\left(\frac{1}{\alpha}\right) & {\rm if} & T\ll\epsilon
\end{array}\right..
\end{equation}
The origin of the additional logarithm is the singular phase space
in collinear scattering processes, an effect that does not enter the
transport rate because of the forward scattering kernel ${\cal K}_{{\rm tr}}$.

At lowest energies $\epsilon\ll\alpha^{2}T$, all scales behave the
same. For $\epsilon\sim\alpha T$ , $\tau_{q}^{-1}\sim\tau_{{\rm tr}}^{-1}\sim\frac{1}{N}\alpha T\gg\tau_{E}^{-1}\sim\frac{1}{N}\alpha^{3/2}\log\alpha^{-1}T$
implying that energy relaxation is the slowest process. For $\epsilon\sim T$,
the dephasing rate is the largest scale $\tau_{q}^{-1}\sim\frac{1}{N}\alpha T\gg\tau_{E}^{-1}\sim\frac{1}{N}\alpha^{2}T\log\frac{1}{\alpha}\gg\tau_{{\rm tr}}^{-1}\sim\frac{1}{N}\alpha^{2}T$.
A representation of these scattering rates as function of energy is
depicted in Fig. \ref{fig:Qualitative-representation-the}.

\section{The homogeneous Bethe-Salpeter equation}\label{TheBSGlDetails}

\label{app:homo-bse}  For further analysis, we express the homogeneous
Bethe-Salpeter equation (see. Eq. \ref{eq:homo-bse}) in terms of
dimensionless variables $K=\frac{v|\mathbf{k}|}{2k_{B}T}$ and replace
the angle integrations by an additional momentum integration. We obtain
the homogeneous integral equation 
\begin{equation}
\lambda\, f(\omega,K)=\frac{4}{N}\frac{2\pi k_{B}T}{\hbar}\,\int_{0}^{\infty}\frac{K'dK'}{2\pi}\mathcal{M}(K,K')\, f(\omega,K')\label{eq:integral-equation-2}
\end{equation}
with $\mathcal{M}=\mathcal{M}_{+}+\mathcal{M}_{-}$. The one-rung
contributions (superscript $\left(1\right)$) are given by 
\begin{eqnarray}
\mathcal{M}_{+}^{(1)}\left(K,K'\right) & = & \frac{2}{KK'}\int_{|K-K'|}^{K+K'}\frac{QdQ}{2\pi}\frac{\sqrt{\left(K+K'\right)^{2}-Q^{2}}}{\sqrt{Q^{2}-\left(K-K'\right)^{2}}}\frac{\text{Im}\mathcal{D}_{ul}^{R}\left(|K-K'|,Q\right)}{\sinh\left(|K-K'|\right)},\\
\mathcal{M}_{-}^{(1)}\left(K,K'\right) & = & \frac{2}{KK'}\int_{|K-K'|}^{K+K'}\frac{QdQ}{2\pi}\,\sqrt{\frac{Q^{2}-\left(K-K'\right)^{2}}{\left(K+K'\right)^{2}-Q^{2}}}\frac{\text{Im}\mathcal{D}_{ul}^{R}\left(K+K',Q\right)}{\sinh\left(K+K'\right)}
\end{eqnarray}
where we introduce the dimensionless imaginary part of the bosonic
propagator as 
\begin{equation}
\text{Im}\mathcal{D}_{ul}^{R}(x,y)=\frac{\left(\frac{\alpha}{2}\right)^{2}\mathcal{I}_{F}\left(x,y\right)}{\left(Q+\frac{\alpha\mathcal{I}_{G}\left(x,y\right)}{2}\right)^{2}+\left(\frac{\alpha\mathcal{I}_{F}\left(x,y\right)}{2}\right)^{2}}.\label{eq:ImDr}
\end{equation}
The dimensionless functions $\mathcal{I}_{F}$ and $\mathcal{I}_{G}$
are defined via the real and imaginary part of the polarization operator
$\text{Im}\Pi^{R}\left(\omega,\mathbf{q}\right)=\frac{k_{B}T}{2\pi}\mathcal{I}_{F}(\frac{\omega}{2k_{B}T},\frac{v|\mathbf{q}|}{2k_{B}T})$
and $\text{Re}\Pi^{R}(\omega,\mathbf{q})=\frac{k_{B}T}{2\pi}\mathcal{I}_{G}(\frac{\omega}{2k_{B}T},\frac{v|\mathbf{q}|}{2k_{B}T})$,
respectively. Their explicit expressions are 
\begin{eqnarray}
\mathcal{I}_{F}\left(x,y\right) & = & \frac{\sinh x}{\sqrt{|x^{2}-y^{2}|}}\begin{cases}
\int_{y}^{\infty}d\xi\frac{\sqrt{\xi^{2}-y^{2}}}{\cosh x+\cosh\xi} & \text{for }y>|x|\\
\int_{0}^{y}d\eta\frac{\sqrt{y^{2}-\eta^{2}}}{\cosh x+\cosh\eta} & \text{for }|x|>y
\end{cases},\\
\mathcal{I}_{G}\left(x,y\right) & = & -\frac{2}{\pi}\int_{y}^{\infty}d\xi\int_{0}^{y}d\eta\left(\frac{\sqrt{\xi^{2}-y^{2}}}{\sqrt{y^{2}-\eta^{2}}}\frac{\eta}{x^{2}-\eta^{2}}\frac{\sinh\eta}{\cosh\eta+\cosh\xi}+\frac{\sqrt{y^{2}-\eta^{2}}}{\sqrt{\xi^{2}-y^{2}}}\frac{\xi}{x^{2}-\xi^{2}}\frac{\sinh\xi}{\cosh\eta+\cosh\xi}\right).
\end{eqnarray}

The two-rung contributions (superscript $\left(2\right)$) are given
by 
\begin{multline}
\mathcal{M}_{+}^{(2)}\left(K,K'\right)=\frac{4\pi}{KK'}\int\frac{\tilde{K}d\tilde{K}}{2\pi}\\
\times\bigg(\int_{\text{max}\left[|K-\tilde{K}|+K,|K'-\tilde{K}|+K'\right]}^{\text{min}\left[2K+\tilde{K},2K'+\tilde{K}\right]}\frac{dQ}{2\pi}\sqrt{\frac{\tilde{K}^{2}-\left(2K-Q\right)^{2}}{Q^{2}-\tilde{K}^{2}}}\sqrt{\frac{\tilde{K}^{2}-\left(2K'-Q\right)^{2}}{Q^{2}-\tilde{K}^{2}}}\frac{|\mathcal{D}(Q,\tilde{K})|^{2}}{\cosh\left(K-Q\right)\cosh\left(K'-Q\right)}\\
+\int_{\text{max}\left[|K-\tilde{K}|-K,|K'-\tilde{K}|-K'\right]}^{\tilde{K}}\frac{dQ}{2\pi}\sqrt{\frac{\left(2K+Q\right)^{2}-\tilde{K}^{2}}{\tilde{K}^{2}-Q^{2}}}\sqrt{\frac{\left(2K'+Q\right)^{2}-\tilde{K}^{2}}{\tilde{K}^{2}-Q^{2}}}\frac{|\mathcal{D}(Q,\tilde{K})|^{2}}{\cosh\left(K+Q\right)\cosh\left(K'+Q\right)}\bigg)
\end{multline}
and 
\begin{multline}
\mathcal{M}_{-}^{(2)}\left(K,K'\right)=\frac{4\pi}{KK'}\int\frac{\tilde{K}d\tilde{K}}{2\pi}\\
\times\int_{\text{min}\left[|K-\tilde{K}|-K,K'-|K'-\tilde{K}|\right]}^{\tilde{K}}\frac{dQ}{2\pi}\sqrt{\frac{\left(2K+Q\right)^{2}-\tilde{K}^{2}}{\tilde{K}^{2}-Q^{2}}}\sqrt{\frac{\left(2K'-Q\right)^{2}-\tilde{K}^{2}}{\tilde{K}^{2}-Q^{2}}}\frac{|\mathcal{D}(Q,\tilde{K})|^{2}}{\cosh\left(K+Q\right)\cosh\left(K'-Q\right)}
\end{multline}
where 
\begin{equation}
|\mathcal{D}(Q,\tilde{K})|^{2}=\frac{\left(\frac{\alpha}{2}\right)^{2}}{\left(Q+\frac{\alpha\mathcal{I}_{G}\left(x,y\right)}{2}\right)^{2}+\left(\frac{\alpha\mathcal{I}_{F}\left(x,y\right)}{2}\right)^{2}}.
\end{equation}

\section{Numerical procedure }\label{NumericalProcedure}

The integral equation \ref{eq:integral-equation-2} is solved numerically
by discretizing the area of integration. We use a homogeneous grid
with up to $2^{9}\times2^{9}$ grid points in $K\times K'$-space
and diagonalize the obtained matrix. We compare our results to results
obtained by solving the integral equation recursively for the one-rung
$\mathcal{M}_{+}^{(1)}$-contribution only where the kernel is evaluated
analytically. In this case, the one-dimensional $K$-domain is discretized
using $10^{4}$ grid points. It turns out that this process contributes
predominantly to $\lambda_{L}$ and serves as a lower bound on the
exponent for weak coupling.

\section{Lyapunov spectrum}\label{LyapunovSpectr}

\begin{figure}[h]
\centering{}\includegraphics[width=0.4\columnwidth]{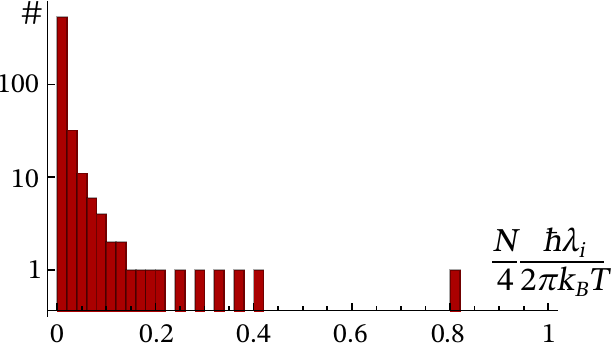}\protect\caption{\label{fig:Spectrum-of-Lyapunov-1}Spectrum of Lyapunov exponents
$\{\lambda_{i}\}$ in the strong coupling regime $\alpha\gg1$. Lyapunov
exponents are binned in intervals of length $0.02$.}
\end{figure}

The spectrum of exponents for a specific coupling is depicted in Fig.
\ref{fig:Spectrum-of-Lyapunov-1}. It is qualitatively the same for
all couplings. We observe that the largest eigenvalue ($\lambda_{L}$)
is well separated from the next-to-largest eigenvalue. This justifies
the discussion about one specific Lyapunov exponent $\lambda_{L}$.

\end{document}